\documentclass[reprint,superscriptaddress,amsmath,amssymb,aps,pra, floatfix]{revtex4-2}

\usepackage[main=english]{babel}

\newcommand{\mycomment}[1]{}
\usepackage{graphicx}
\usepackage{dcolumn}
\usepackage{bm}
\usepackage[T1]{fontenc}
\usepackage{geometry}
\begin{document}

\preprint{APS/123-QED}

\title{A non-reciprocal model for morphogenesis in symbiosis}

\author{Maitane Muñoz-Basagoiti}
\affiliation{
 Institute of Science and Technology Austria, Am Campus, 3400 Klosterneuburg, Austria
 }
\author{Michael Wassermair}
\affiliation{
 Institute of Science and Technology Austria, Am Campus, 3400 Klosterneuburg, Austria
 }
\author{Miguel Amaral}
\affiliation{
 Institute of Science and Technology Austria, Am Campus, 3400 Klosterneuburg, Austria
 }
\author{Buzz Baum}
\affiliation{
MRC Laboratory of Molecular Biology, Cambridge, UK}
\author{An\dj ela Šarić}
\email{andela.saric@ista.ac.at}
\affiliation{
 Institute of Science and Technology Austria, Am Campus, 3400 Klosterneuburg, Austria
 }

%\date{\today}

\begin{abstract}
The shape of a cell influences, and it is influenced by its interactions with its neighbours. Here, we introduce a coarse-grained computational model of non-reciprocal interactions between single-cell organisms to study emergent morphologies during symbiotic association. We show that the cell membrane can be remodelled into branched protrusions, invaginations, transient blebs and other dynamical morphologies that depend on the number of interacting partners, the asymmetry, and the magnitude of partnership activity. Our model finds a dynamical feedback between the local deformation of the membrane and its driving force, leading to membrane morphologies not reported in reciprocal systems with constant activity. 
\end{abstract}

\maketitle

Metabolic exchanges constitute the basis of symbiotic associations~\cite{kost_metabolic_2023}.
These chemical exchanges impact the fitness of the interacting organisms, and dictate the stability of the associations, from microbial communities~\cite{faust2012microbial, seth2014nutrient} to host-microbe symbioses~\cite{visick_lasting_2021, maslowski2019metabolism, junker2025dynamic}.
Beyond the identity of the chemicals exchanged, interactions between symbiotic partners depend on their shape, size, surface area and volume~\cite{cornejo2024metabolic, biquand2017acceptable, nardon_morphological_2002}. These geometrical constraints can drive self-organised patterns in populations of interacting cells~\cite{smith2017cell}, and impact the global structure of microbial communities~\cite{storck2014variable}, demonstrating that cell morphology plays a fundamental role in metabolic interactions~\cite{Imachi2025.02.26.640444}.

Since cells are able to dynamically remodel their shape, symbiotic interactions can also drive the emergence of morphologies that optimise metabolic exchanges between partner cells. 
Examples include the tunnelling nanotubes that connect bacteria~\cite{baidya_bacterial_2018}, the cytoplasmic invaginations that plant root cells develop to host fungi~\cite{parniske_arbuscular_2008}, and the highly branched villous network developed between the placenta and fetus during pregnancy~\cite{cross_branching_2006}. 
These deformations involve numerous molecular players~\cite{zhang2025networks}, and can span the lifetime of an organism~\cite{junker2025dynamic, leander_symbiotic_2004, visick_lasting_2021}, or even take place in generational timescales~\cite{margulis_symbiosis_1971, Baum2014}. Therefore, predicting the cellular morphologies that emerge from symbiotic interactions, and establishing a link between the morphology of cells and their interactions in symbiosis remains a challenge.
 
In symbiosis, the metabolites that one cell shares with another are, in general, different from what it receives in return~\cite{jahn_nanoarchaeum_2008, kost_metabolic_2023}. Consequently, the effective interactions between symbiotic partners need not be reciprocal. Non-reciprocal interactions can naturally arise when coarse-graining the degrees of freedom of non-equilibrium systems~\cite{fruchart_non-reciprocal_2021, klapp_non-reciprocal_2023}. These interactions have recently provided a powerful framework to describe active matter phenomena, ranging from the dynamics of catalytically coated colloids~\cite{Soto2014} and oil droplet dynamics in mycellar solutions~\cite{meredith_predatorprey_2020}, to pattern formation~\cite{saha_scalar_2020, brauns_nonreciprocal_2024}, multifarious self-assembly~\cite{osat_non-reciprocal_2023}, enzymatic self-organization~\cite{ouazan-reboul_self-organization_2023}, and vision-cone interactions~\cite{loos_long-range_2023}, among others. These studies, however, consider interacting agents whose morphology is fixed or plays little role in the dynamics of the system. Extending non-reciprocal interactions to deformable cells provides a minimal framework to start investigating how metabolic interactions can drive changes in cell shape, and how the resulting shapes can, in turn, modulate these interactions during symbiosis.

\begin{figure*}[ht!]
    \centering
    \includegraphics[width=\linewidth]{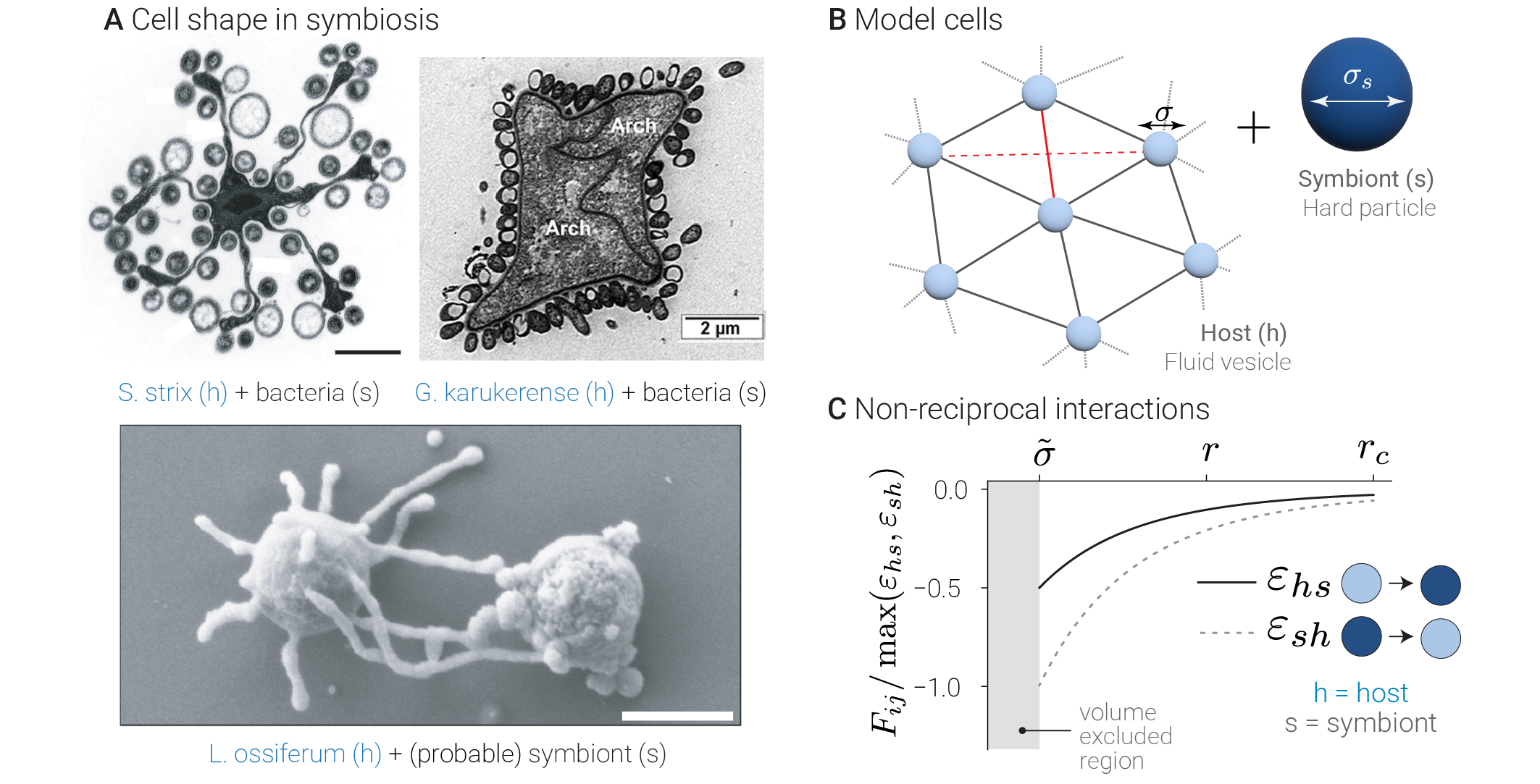}
    \caption{\textbf{What is the interplay between symbiotic interactions and cell shape?} \textbf{A.} Morphologies adopted by microorganisms in symbiotic association. \textit{S. strix} image reprinted with permission from~\cite{leander_symbiotic_2004}; \textit{G. karukense} image reprinted with permission from~\cite{muller_first_2010}. \textit{L. ossiferum} image reprinted with permission from~\cite{RodriguesOliveira2023}. 
    Arrows in the original \textit{S. strix} image have been removed to aid in the visualization. \textbf{B-C.} Ingredients of the coarse-grained computational model. B. We model the host organism as a dynamically triangulated network. The vertices of the network interact with symbiont partners, which are modelled as hard spheres with diameter $\sigma_s>\sigma$, where $\sigma$ is the effective diameter of the vertices in the triangulated network. The fluidity of the plasma membrane in the host organism is achieved by periodically updating the connectivity of the network through bond swaps (red; SI). C. Host and symbiont particles interact through non-reciprocal attractive forces, where $\tilde{\sigma} = (\sigma+\sigma_s)/2$ and $r_c$ is the interaction range. Reciprocal volume exclusion between host and symbiont particles prevents overlaps when $r\leq \tilde{\sigma}$ (grey area).}
    \label{fig:fig1}
\end{figure*}

In this manuscript, we introduce a coarse-grained computational model  based on pairwise non-reciprocal interactions between deformable model cells, to study the ensemble of morphologies that can emerge from cells engaged in symbiosis.
Despite the simplicity of the model, we observe the emergence of diverse shapes that resemble cell morphologies seen in ecological partnerships (Fig.~\ref{fig:fig1}A), including the formation of long membrane protrusions and membrane invaginations. Crucially, these deformations are induced by activity that results from the interaction between otherwise passive model cells. 
Furthermore, we show that the non-reciprocity in the interactions naturally leads to a feedback between the deformation of the cells and the activity available to generate those deformations. These effects result in a variety of collective effects which have not been previously described in models that use motile filaments and other self-propelled agents to deform membranes~\cite{Vutukuri2020, Iyer2022, Iyer2023}. 
Together, our results suggest that the feedback between the shape of a cell and its metabolic interactions with symbiotic partners could sustain morphologies that differ from those generated by the cell and its machinery in isolation, a feature already recognised for multicellular symbiotic organisms~\cite{pichler2023build, visick_lasting_2021, junker2025dynamic}.

\vspace{1mm}
\noindent \textbf{Computational model ---}
We consider a deformable host cell and its non-deformable symbiotic partners, modelled as a fluid lipid vesicle and volume-excluded spheres, respectively (Fig~\ref{fig:fig1}B).
To simulate the vesicle, we use a dynamically triangulated network of beads with diameter $\sigma$ (simulation unit of length) connected by bonds with constrained lengths~\cite{Siggel2022, munoz-basagoiti_tutorial_2025}. 
The vesicle Hamiltonian contains energy contributions from bending, area stretching and volume fluctuations.
In particular, the total cell surface area and enclosed vesicle volume are allowed to fluctuate around fixed values using quadratic energy constraints. 
Membrane fluidity is achieved by regularly updating the connectivity of the network through Monte Carlo bond-swaps~(Fig~\ref{fig:fig1}B).
 Further details are provided in the Supplementary Information (SI).

\newpage
Symbiont particles have diameter $\sigma_s$, and interact with particles on the vertices of the vesicle network through non-reciprocal pairwise forces (Fig.~\ref{fig:fig1}C).
We consider that a particle $k$ from species $i$, where $i$ can refer to host beads~($h$) or symbiont particles~($s$), exerts a force on a particle $l$ of the other species $j$, with $j \in \{h, s\}$ and $i\neq j$, equal to 
\begin{equation}\label{f1}
    \vec{F}_{k,i\rightarrow l,j} = -\phi \frac{\varepsilon_{ij}}{r^{\gamma}} \hat{r}_{kl} \quad \text{if } \quad {r<r_c}
\end{equation}
where $r$ is the distance between particles, $\hat{r}_{kl} = \hat{r}_{l} - \hat{r}_{k}$, with $\hat{r}_{k}$ the unit position vector for particle $k$, $\gamma$ is an exponent that controls the range of the interaction, $\phi$ is a constant that sets the scale of the force, and $r_c$ is the range of the interaction, which we set $r_c = 1.5~\tilde{\sigma}$, where $\tilde{\sigma} = (\sigma + \sigma_s)/2$. 
Additionally, all the particles in the system have a volume-excluded core to prevent overlaps. We use Molecular Dynamics simulations coupled to a Langevin thermostat to integrate the dynamics of the particles in the system (SI).

To model the biochemical complexity of metabolic exchanges, which are generally asymmetric both in the nature of the chemicals exchanged, as well as their quantity~\cite{kost_metabolic_2023, jahn_nanoarchaeum_2008}, we allow the force that each particle species exerts on the other species to break action-reaction symmetry, i.e., $\vec{F}_{sh} \neq -\vec{F}_{hs}$. This is achieved by choosing the strength of the interaction such that $\varepsilon_{hs} \neq \varepsilon_{sh}$, where $\varepsilon_{hs}$ quantifies the magnitude of force that the host vesicle exerts on the symbiont, and conversely, $\varepsilon_{sh}$ the magnitude of the force that a symbiont exerts on the host. 
We define the parameter $\Delta \varepsilon = \varepsilon_{sh} - \varepsilon_{hs} $ to characterize the deviation of the forces defined in Eq.~(\ref{f1}) from the reciprocal case, which thus corresponds to $\Delta \varepsilon = 0$. We note that an alternative definition for the interaction parameters could have been to choose $\varepsilon_{ij}$ such that $\varepsilon_{sh} = \varepsilon_R + \varepsilon_{NR}/2$ and $\varepsilon_{hs} = \varepsilon_R - \varepsilon_{NR}/2$, where $\varepsilon_R$ and $\varepsilon_{NR}$ correspond to the magnitude of the reciprocal (R) and non-reciprocal (NR) part of the interactions between model cells. This would have resulted in $\Delta \varepsilon = \varepsilon_{NR}$. Instead, in what follows we will change $\varepsilon_{hs}$ and $\varepsilon_{sh}$ independently, which highlight the individuality of the metabolic exchanges.

\begin{figure*}[ht!]
    \centering\includegraphics[width=\linewidth]{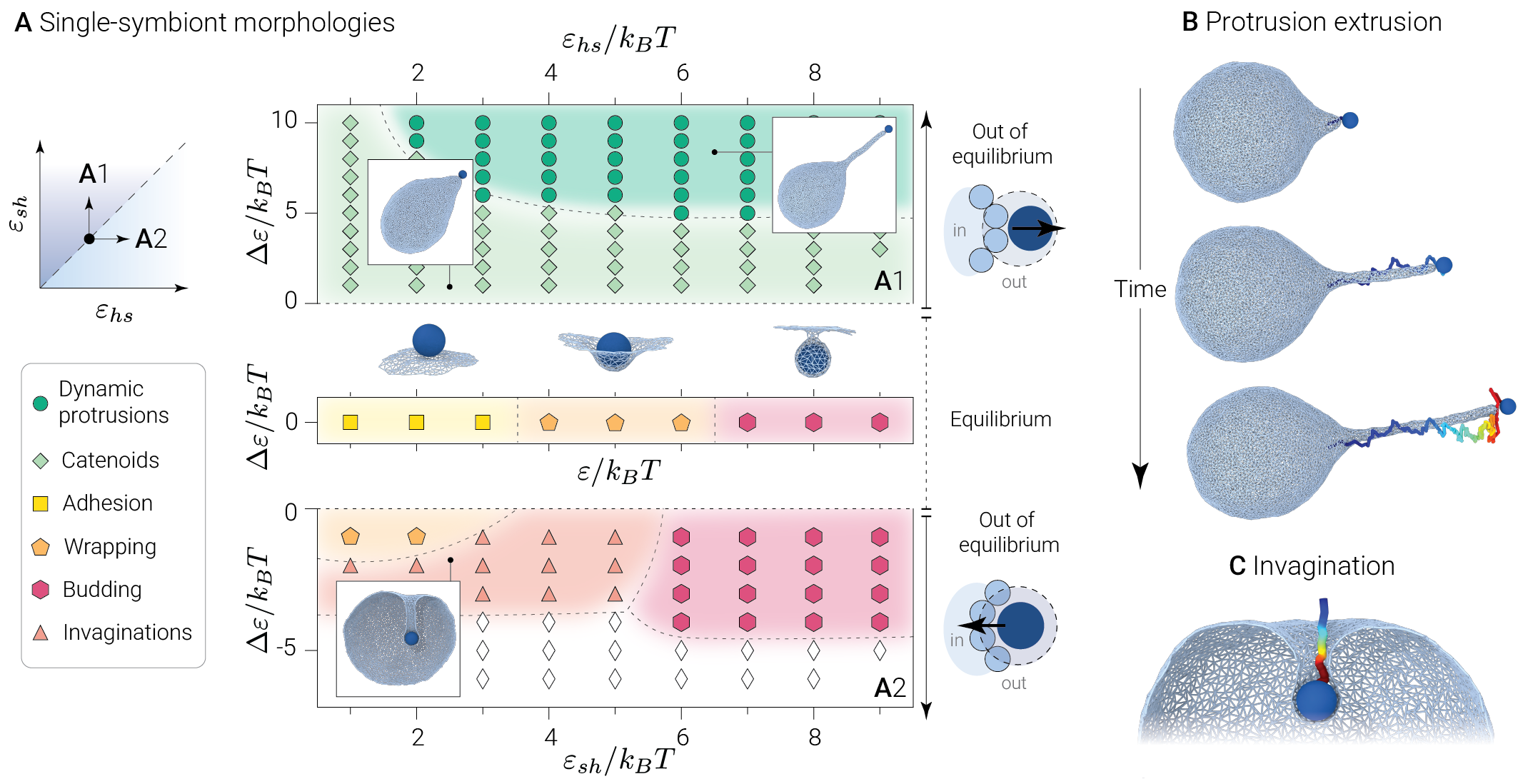}
    \caption{\textbf{Morphologies generated when a deformable model cell interacts with a single symbiont particle.} ~~\textbf{A.} Ensemble of morphologies that emerge when the host model cell ($h$) and a single symbiont particle ($s$) interact via (non-)reciprocal interactions, where $\Delta \varepsilon = \varepsilon_{sh} - \varepsilon_{hs}$ characterises the asymmetry in the interactions (reciprocal for $\Delta \varepsilon = 0$, and non-reciprocal for $\Delta \varepsilon> 0 $ and $\Delta \varepsilon < 0$). The quadrants in the morphology diagrams correspond to the upper (A1) and lower (A2) triangles of the parameter space spanned by $(\varepsilon_{sh}, \varepsilon_{hs})$. The symbiont particle has diameter with $\sigma_s/\sigma = 5$ and its size relative to the model cell is $\sigma_h/\sigma_s \approx 6$. White data points indicate the region in parameter space where the membrane is expected to change topology and cannot be explored in the triangulated model. Simulation parameters and morphology identification criteria are detailed in the SI; dashed lines are guides for the eye, and each point corresponds to 10 simulation replicas. \textbf{B.} Growth of a membrane protrusion. \textbf{C}. Invagination of the model cell membrane (SI Movies 1-3). Trajectories show the position of the symbiont in time, from blue to red.}
    \label{fig:fig2}
\end{figure*}

We choose the parameters $\gamma$ and $\phi$ such that we can establish a connection between Eq.~(\ref{f1}) and the force derived from the canonical Lennard-Jones potential, which corresponds to $\gamma = 7$ and $\phi = 6/\tilde{\sigma}$.
This allows us to compare the outcome of our out-of-equilibrium simulations to the equilibrium morphologies that result from the canonical passive adhesion of particles and organisms to the fluid membrane of the host model cell~\cite{deserno_elastic_2004, saric_self-assembly_2013, Saric2012}. 
Alternative choices of $\gamma$ would allow to capture other classes of non-reciprocal interactions studied in soft matter systems. In particular, the case of $\gamma = 2$ can be mapped
to the phoretic activity that catalytically coated colloids acquire upon hydrodynamic interaction through chemical exchanges in the far-field approximation, where $\varepsilon$ would account for the product of the activity and mobility of the interacting colloids~\cite{Soto2014}.
We set $\gamma = 7$ throughout the paper, and check that the choice of exponent does not alter qualitatively the emerging morphologies (SI Fig.~1). %This suggests that our results can account for metabolically-coupled symbiotic partners, i.e., intercellular communication driven by the exchange of chemicals.

\vspace{1mm}
\noindent\textbf{Results ---} Symbiotic interactions between organisms can be favourable or unfavourable ~\cite{kost_metabolic_2023}. We use the sign of $\varepsilon_{ij}$ in Eq.~(\ref{f1}) to encode for the type of association between organisms, and consider that favourable interactions drive organisms together (attractive forces, $\varepsilon_{ij}>0$), while unfavourable interactions bring them apart (repulsive forces, $\varepsilon_{ij}<0$).
We focus on investigating the case where both forces are attractive, i.e., $\varepsilon_{ij}>0~\forall i, j\in s, h$, $i\neq j$,  which ensures that the model cells generally remain within interaction range throughout the simulation. This choice can be interpreted as a model for mutualism~\cite{levin_princeton_2012}, where cells engage in an association that benefits both parties. 

\vspace{1mm}
\noindent\textit{Morphologies in interaction with single symbiont.} To explore the generation of cell shape in our model, we first place a single symbiont within interaction range of the host vesicle, and let the organisms interact through the pairwise forces in Eq.~(\ref{f1}). We choose our model cells to have relative diameters $\sigma_h / \sigma_s\approx 6$, where $\sigma_h$ is the diameter of the model deformable cell. As the interaction is characterized by both $\varepsilon_{sh}$ and $\varepsilon_{hs}$, we fix a $\varepsilon_{ij}$ parameter as a reference, and vary the remaining $\varepsilon_{ji}$ to map out the parameter space. The emerging morphologies are summarized in Fig.~\ref{fig:fig2}.

\begin{figure*}[t!]
    \centering
    \includegraphics[width=\linewidth]{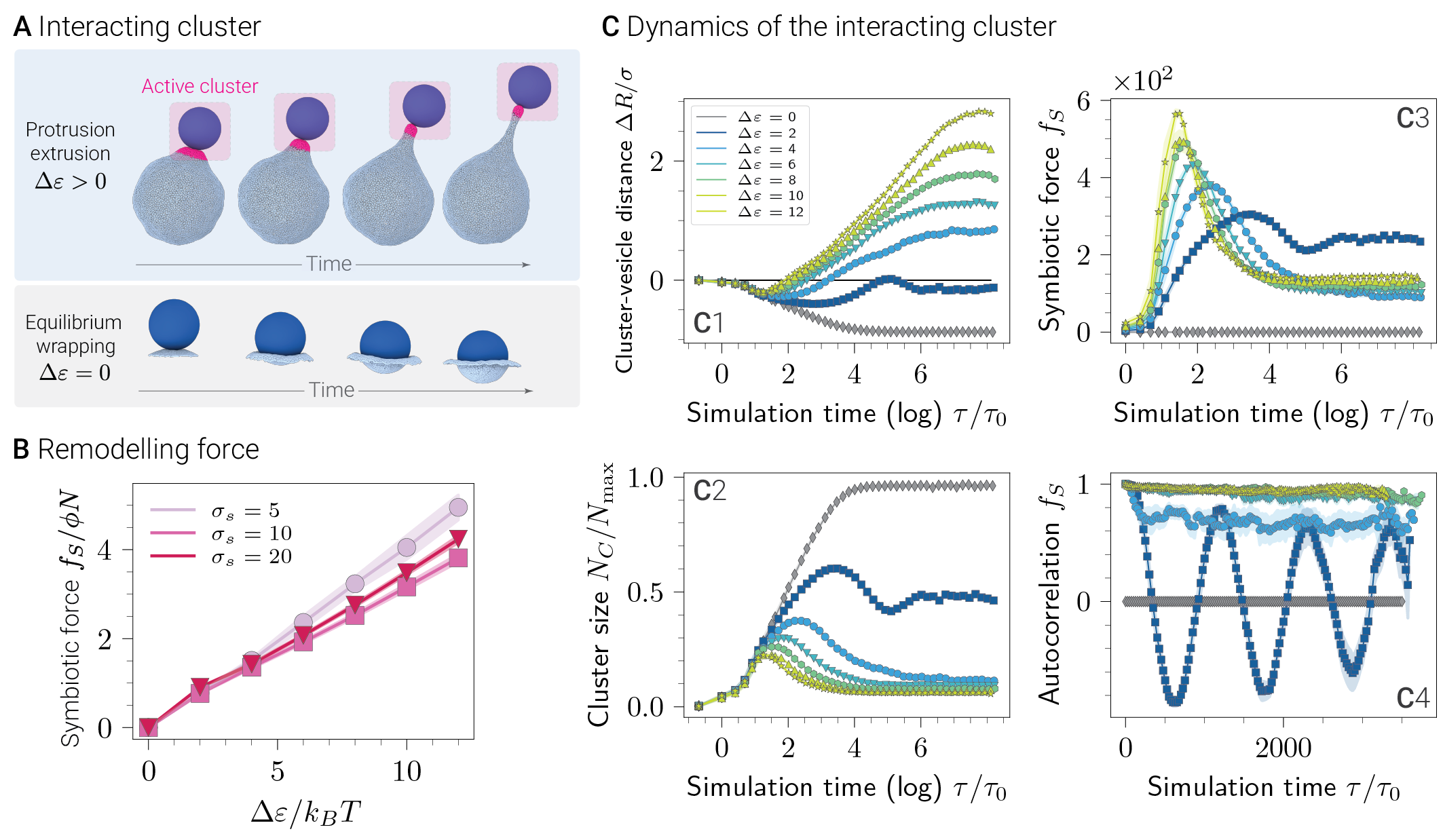}
    \caption{\textbf{Dynamical properties of the interaction site.} \textbf{A.} Clusters of interacting particles and model cell deformations when host and symbiont particles interact reciprocally (bottom; only the interacting cluster is shown) and non-reciprocally (top; interacting cluster is shaded in pink). The symbiont particle has a diameter $\sigma_S = 20~\sigma$. The patch of membrane within the interacting cluster is highlighted in bright pink for $\Delta \varepsilon$. \textbf{B.}~Steady-state symbiotic force $f_{S}$ available to remodel the model cell, normalized by the size of the active cluster $N_C$ and $\phi$, as a function of $\Delta \varepsilon$ for three symbiont sizes. \textbf{C.} Position of the interacting cluster, quantified as its distance to the centre of mass of the deformable model cell, and normalized by the vesicle radius (C1), size of the interacting cluster $N$ normalized by the equilibrium cluster size $N_{\max}$ (C2), magnitude of the symbiotic force (C3) and its autocorrelation function (C4) as a function of time. We fix $\varepsilon_{hs} = 10 ~k_B T$, and change $\varepsilon_{sh}$. The grey data points show the reciprocal case $\Delta \varepsilon = 0$, while coloured data correspond to $\Delta \varepsilon >0$. Data in C corresponds to a symbiont with $\sigma_s = 20\sigma$; $n_{\text{replicas}}>10$.}
    \label{fig:fig3}
\end{figure*}

When interactions are reciprocal, that is, at equilibrium ($\Delta \varepsilon = 0$ line in Fig.~\ref{fig:fig2}A), we observe the progressive wrapping of the symbiont particle with increasing interaction strength $\varepsilon = \varepsilon_{hs} = \varepsilon_{sh}$, as expected for passive nanoparticle wrapping~\cite{deserno_elastic_2004} (SI Fig.~2).
Instead, the region corresponding to $\Delta \varepsilon >0$ ($\varepsilon_{sh} > \varepsilon_{hs}$) is dominated by the emergence of protrusions (Fig.~\ref{fig:fig2}B), which form when the interaction between symbiotic particles satisfies $\Delta \varepsilon/k_B T \gtrsim 5$.
Conversely, in the $\Delta \varepsilon < 0$ region ($\varepsilon_{sh} < \varepsilon_{hs}$), the symbiont particle can end up partially wrapped, invaginated (Fig.~\ref{fig:fig2}C), or fully budded by the host vesicle, depending on the value of $\varepsilon_{sh}$.

The deformations reported in Fig.~\ref{fig:fig2} do not always reach a steady-state. When $\Delta \varepsilon>0$, we frequently observe symbiont particles move away from the range of interaction with the deformable vesicle. If this happens, the emergent protrusion retracts and the model cell vesicle recovers its equilibrium shape (SI Movie~1). In this region of the parameter space, the lifetime of the interaction depends on $\varepsilon_{hs}$. In particular, when $\varepsilon_{hs}/k_B T = 2$, the lifetime of the interaction is so short that the model cells stop interacting before the protrusion can grow. Likewise, the invaginations that emerge for $\Delta \varepsilon<0$ are also transient: the opening of the invagination moves, and the symbiont particle ends up fully wrapped. See criteria for protrusion and invagination detection in the SI.

\newpage
To verify that our simulation results do not depend on the choice of membrane model, we have reproduced the morphology diagram in Fig.~\ref{fig:fig3}A using a different model: the self-assembled one-particle thick membrane model that can undergo fission (SI Fig.~3)~\cite{Yuan2010}. 
The white diamond symbols in Fig.~\ref{fig:fig2}A correspond to simulations where the self-assembled membrane ruptures upon interaction with the symbiont, and therefore, constitute the region of the parameter space which we cannot properly sample using a triangulated membrane model that cannot change topology. The fixed topology of the membrane model also explains the membrane neck of budded configurations. We next discuss the origin of the shape-generating activity in the system, and its coupling to the deformation of the model cell membrane.

\vspace{1mm}
\noindent\textit{Dynamics of the interaction site.} A symbiont and the particles on the vesicle network within interaction range form a cluster of interacting particles (Fig.~\ref{fig:fig3}A).
When interactions are reciprocal ($\Delta \varepsilon = 0$), and in the absence of external forces, the net force on the Center Of Mass (COM) of such cluster is zero due to action-reaction symmetry.
However, when the action-reaction principle is no longer satisfied ($\Delta \varepsilon \neq 0 $), the force on the COM of the cluster does not cancel out in general except for fully symmetric clusters. 
Such force $f_S$, which only emerges when the host model cell and the symbiont particle come into contact, and which acts on the site of interaction between model cells, is at the origin of the cell deformations in our model.

\begin{figure*}[ht!]
    \centering\includegraphics[width=\linewidth]{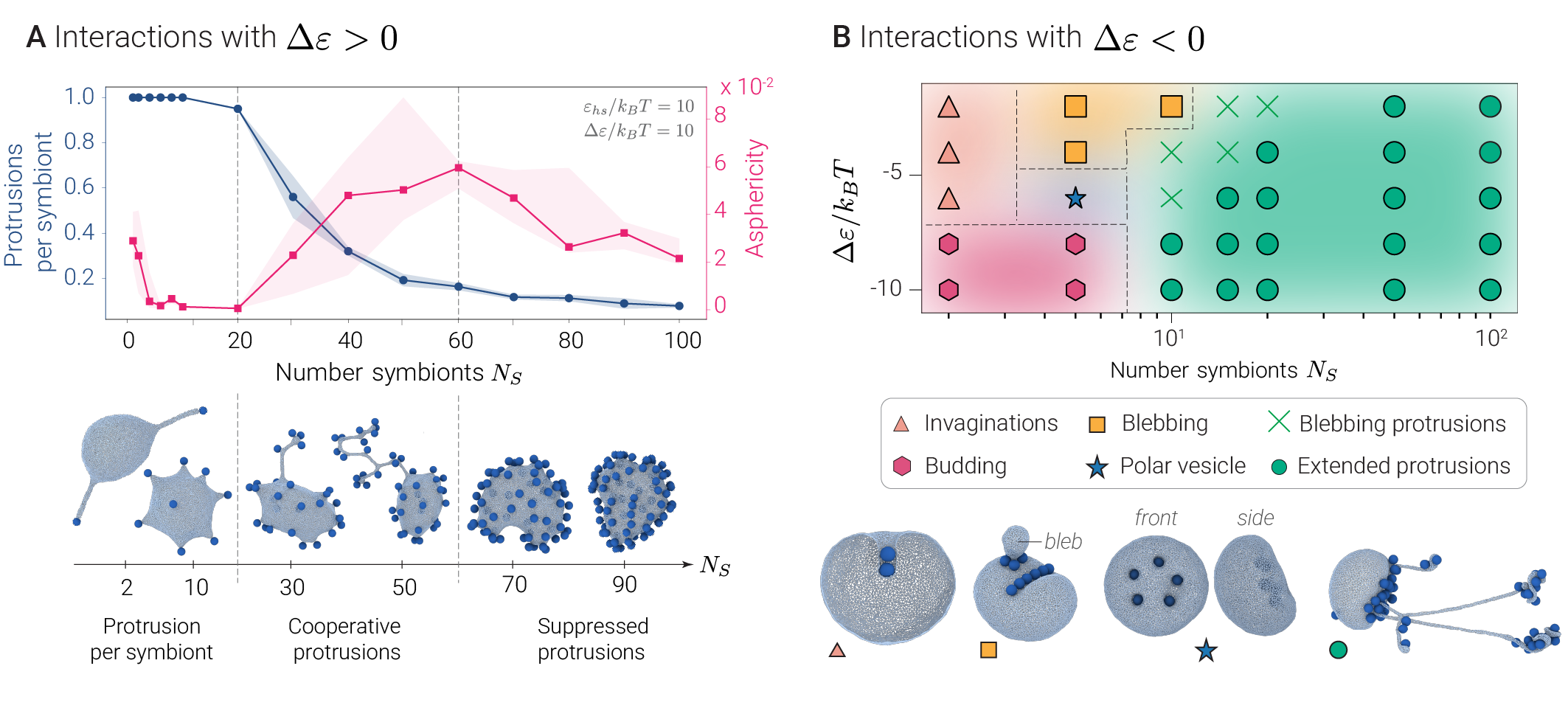}
    \caption{\textbf{Morphologies generated when a deformable model cell interacts with multiple symbionts}. \\ \textbf{A.} Maximum number of protrusions (blue) and model cell  asphericity (pink) as a function of the number of interacting symbionts $N_S$, for $\Delta \varepsilon/k_BT = 10$, and $\varepsilon_{hs}/k_B T= 20$ and $\sigma_s = 5~\sigma$. Protrusions form independently at low $N_S$. At intermediate $N_S$, the growth of a protrusion requires the clustering of several symbionts. At hight $N_S$ protrusions are suppressed. See SI for the criteria to identify the protrusions; $n_{\text{Replica}} = 3$. \textbf{B.} Emerging morphologies when $\Delta \varepsilon<0$ and $\varepsilon_{sh}/k_BT = 2$. As a function of interacting symbionts $N_S$ and $\Delta \varepsilon$, we observe symbiont budding and invaginations, the cyclic formation of blebs on the vesicle surface, polar symbiont distributions on the surface of the host cell, and membrane protrusions. Green crosses indicate short and irregular protrusions (SI Fig.~6), while green circles correspond to protrusions extruded as a result of asymmetric-coating vesicles (depicted). Dashed lines are guides for the eye; $n_{\text{replica}} =3$. Morphologies are identified based on the configuration of the system at the steady-state (SI Fig. 7).}
    \label{fig:fig4}
\end{figure*}

By definition, $f_S$ will point on average towards the symbiont when $\Delta \varepsilon>0$, and towards the membrane beads when $\Delta \varepsilon<0$ (Fig.~\ref{fig:fig2}A), and its magnitude will be directly proportional to $|\Delta \varepsilon|$, since it results from the asymmetry in the interactions (see Eq.~\ref{f1}).
In contrast, the magnitude and directionality of $f_S$ are highly non-trivial, as they depend on the geometry of the interacting particle cluster~\cite{Soto2014}, that is, on the deformation of the model cell vesicle around the symbiont particles (Fig.~\ref{fig:fig3}B).
This however implies that $f_S$ does not necessarily remain constant for fixed $\Delta \varepsilon$, as the overall geometry of the interacting cluster progressive changes during the deformation of the model cell vesicle. Therefore, since the local deformation of the host membrane depends on its interaction with the symbiont, our non-reciprocal model of interacting cells successfully achieves a feedback between the cell shape and symbiotic interactions.

To further illustrate the feedback between cell shape and interactions in the model, we characterize the dynamical properties of the interacting cluster within the $\varepsilon_{hs}/k_BT = 10$ and $\Delta \varepsilon>0$ line, where the interaction results in protrusions (Fig.~\ref{fig:fig3}A).
During the early stage of the interaction, the model organisms attract each other as expected for $\varepsilon_{ij}>0$, which leads to an increase in the size of the interacting cluster (Fig.~\ref{fig:fig3}C1-C2).
This trend monotonically continues for the reciprocal case (gray data) until an equilibrium state is reached, where the cost of bending the membrane and the energy gained from interacting with the symbiont balance each other.
However, when $\Delta \varepsilon > 0$, the curves in Fig.~\ref{fig:fig3}C are non-monotonic. In such case,
the increasing size of the interacting cluster drives an increase in the remodelling force $f_S$ (SI Fig. 4). When $f_S$ is large enough, the interacting cluster moves away from the vesicle surface, generating a protrusion.

Extruding a protrusion from a fluid membrane is a well studied process. It relies on applying a force $f_P\sim \sqrt{\kappa \Sigma}$ on the membrane to generate the deformation, where $\kappa$ and $\Sigma$ are the bending modulus and surface tension of the membrane, respectively~\cite{derenyi2002}.
Nonetheless, in our model we do not explicitly pull on the membrane to generate the deformation, and thus, our simulations show that the extrusion of a protrusion is then only possible as long as the directionality of $f_S$ is persistent (Fig.~\ref{fig:fig3}C4), even if its magnitude is above the theoretical threshold $f_S>f_P$ (SI Fig.~5).  

\vspace{1mm}
\noindent \textit{Morphologies in interaction with multiple symbionts.} Symbiotic partnerships at the microscale generally consist of numerous of symbionts interacting with larger host organisms (Fig.~\ref{fig:fig1}A)~\cite{muller_first_2010}. Thus, we next explore the deformations that emerge when a collection of small symbionts interacts with a deformable model cell. In particular, we consider $N_S$ identical and initially homogeneously distributed symbiont particles near the host cell. For simplicity, we set every symbiont particle to interact in the same (non-)reciprocal manner with the host. Symbiont particles will interact with each other through volume exclusion only.

\vspace{1mm}
\noindent \textit{--- Emerging morphologies for $\Delta \varepsilon>0$.} We have shown that when $\Delta \varepsilon>0$ the interaction between a single symbiont particle and a deformable model cell can lead to the extrusion of membrane protrusions (Figs.~\ref{fig:fig2}-\ref{fig:fig3}).
To determine how the simultaneous interaction with multiple symbiont particles impacts these deformations, we pick a point in the morphology diagram deep within the protrusion region (i.e., $\varepsilon_{hs}/k_BT = 20$ and $\Delta \varepsilon/k_BT=10$), and vary the number of symbionts $N_S$ in contact with the membrane. The resulting protrusions are shown in Fig.~\ref{fig:fig4}A.

At low symbiont coverage, protrusion growth is mostly independent of the presence of other symbionts. 
These protrusions however become shorter with increasing $N_S$, as evidenced by the decrease in vesicle asphericity (Fig.~\ref{fig:fig4}A, pink line). This is due to a decrease in the membrane area available to accommodate deformations as $N_S$ grows. At intermediate values of $N_S$, we identify a second regime for protrusion formation shown by the increase in membrane asphericity, which requires symbionts to locally aggregate in order to yield a larger $f_S$, and overcome the barrier for protrusion extrusion.
Interestingly, these protrusions can develop branches, as each symbiont particle in the aggregate will interact with different patches of membrane along the main extrusion.
Finally, at high symbiont coverage, membrane tension is so large that the symbionts cannot overcome the barrier for protrusion formation. Protrusions are then suppressed.

\vspace{2mm}
\noindent \textit{--- Emerging morphologies for $\Delta \varepsilon<0$.} 
Results in Fig.~\ref{fig:fig2} show that the deformations that emerge for $\Delta \varepsilon < 0$ greatly depend on the parameter $\varepsilon_{sh}$, which dictates the extent of the membrane wrapping around the symbiont, and consequently, the remodelling force available to deform the model cell (Fig.~\ref{fig:fig3}C).
Here focus on the deformations that emerge when symbiont particles weakly interact with the model cell, namely, $\varepsilon_{sh}/k_B T = 2$, since it allows for a larger range of non-symmetric (non-budded) interacting clusters, and consequently, a more diverse set of morphologies.

The morphology diagram that emerges is very rich in dynamic phases (Fig.~\ref{fig:fig4}B).
We again find that at low symbiont coverage, each symbiont generates its own independent deformation. In such regime, we observe the formation of invaginations which transition into symbiont budding with increasing $|\Delta \varepsilon|$, as in the single-symbiont case.
Nonetheless, as the number of symbionts $N_S$ increases, the surface of the model vesicle starts to couple the dynamics of the group of symbionts. 

Remarkably, at intermediate values of $N_S$ and $\Delta \varepsilon$, the non-reciprocal interaction with the deformable surface of the model cell induces collective dynamics of the group of symbiont particles, which do not interact with each other beyond volume exclusion. These collective effects are best exemplified by cyclic formation of membrane blebs (SI Movie 4), a dynamic morphology that has not previously been reported in the context of self-propelled particles interacting with fluid membranes~\cite{Vutukuri2020, Iyer2022, Iyer2023}.
At comparable symbiont coverage and larger activities, the wrapping of the membrane around each symbiont particle increases, preventing their aggregation.
Nonetheless, symbionts spontaneously end up asymmetrically distributed on the surface of the model cell, forming a Janus-like polar vesicle which displaces ballistically (SI Movie 5). 

Lastly, we also report the formation of two different classes of protrusions which qualitatively from those reported in Fig.~\ref{fig:fig2}. Here symbionts particles localize in ring-like formations at the base of the protrusion instead of its tip (green crosses in Fig.~\ref{fig:fig4}B; SI Fig.~6 and SI Movie 6). Unlike protrusion extrusion by pulling, the emergence of these protrusions is driven by changes in the the surface area and volume of the vesicle. In particular, the excess membrane area generated when the volume of the vesicle decreases due to symbiont wrapping and invaginations is released as a tube, an effect previously studied in Ref.~\cite{bahrami_formation_2017}.
Lastly, protrusions can also form when symmetric distributions of symbionts lock in space portions of a vesicle whose body is being displaced (green circles in Fig.~\ref{fig:fig4}B; SI Movie 7).

\vspace{2mm}
\noindent \textbf{Discussion ---} The shape of a cell plays a fundamental role in the interaction with its neighbouring partners,
which in turn can further induce morphological changes in the cell. Such intercellular exchanges, prominently exemplified by symbiotic associations at the micro-scale, are, in general, not symmetric: what one organism
shares with its partner typically differs from what it may receive in return~\cite{kost_metabolic_2023}. Inspired by the asymmetry of these exchanges, here we have introduced a coarse-grained computational model based on non-reciprocal interactions between deformable cells to study the ensemble of morphologies that can emerge in symbiosis. 

Our model relies on the emergent dynamics of non-reciprocally interacting particle clusters to generate the deformations of the cell (Fig.~\ref{fig:fig3}C). It is thus important to note that it is naturally limited in scope. To keep the model general, we have not explicitly defined the nature of the non-reciprocal interactions between the model cells.
Due to coupled metabolic activity, many different processes can be at play between symbiotic partners. These range from the reception and uptake of symbiont-excreted molecules at the cell surface~\cite{clarke_transport_2014}, to the processing and self-organised response orchestrated by the cell with its internal machinery~\cite{Radler2025.11.30.690169}. Therefore, it is challenging to translate the magnitudes of the interactions in our model to quantitative values associated to cell shape changes, and to define the timescales at which the interactions operate. The steady-states of the morphologies that we report can be interpreted as a subset of cell shapes accessible to metabolically active cells in interaction. In doing so, our model expands the ensemble of shapes reported for fluid membranes out-of-equilibrium~\cite{Vutukuri2020, Iyer2022, Iyer2023}. In the future, it will be interesting to investigate how symbiont particles with more complex shapes, rugged, i.e., particle-decorated surfaces, or even themselves deformable impact the resulting morphologies of the host cell.

Progress in the field of synthetic cells and active matter suggests that our results can be readily tested experimentally~\cite{sharan2023pair, berezney2026active}. The controlled realization of membrane protrusions through non-reciprocal communication could allow the controlled exchange of matter between groups of (reconstituted) and wild-type cells~\cite{ji_self-regulated_2023}. 
We hope our results will motivate experimental efforts towards the generation of out-of-equilibrium cells shapes through asymmetric couplings between deformable synthetic cells and their environment.

\vspace{2mm}
\textbf{Acknowledgements ---} We acknowledge support from the IST-BRIDGE fellowship under the Marie Skłodowska-Curie Grant Agreement No. 101034413 (M.M.-B.), the Allen Distinguished Investigator Award (M.M.-B., B.B. and A. Š.) and the ERC Starting Grant “NEPA” 802960 (A. Š.). We thank Daan Frenkel, Tanniemola Liverpool and Ramin Golestanian for helpful discussions. 

\vspace{2mm}
\textbf{Data availability ---} The simulation and analysis codes to reproduce the results are available as a repository.

\bibliography{bibliography}

\begin{center}
    
SUPPLEMENTARY\\INFORMATION
\end{center}
%\date{\today}

\tableofcontents
\renewcommand{\thefigure}{S\arabic{figure}}
\setcounter{figure}{0}
\section{Simulation details}
\subsection{Fluid membrane model}
We have chosen a membrane model with fixed topology to simulate lipid vesicles in order to avoid dealing with membrane rupture (see section~\ref{yuansection}), which allows us to extrude long protrusions from the host vesicle. Simulations are conducted using the TriLMP software, which is freely available as a repository~\cite{TriLMP}, and it is a coupling of the  TriMem membrane simulation package~\cite{Siggel2022} to LAMMPS~\cite{Thompson2022}. 
TriLMP (TriMem + LAMMPS) enables the simulation of dynamically triangulated networks in LAMMPS by using the efficient parallelization scheme introduced by Siggel \textit{et al.}~\cite{Siggel2022}. The fluid membrane Hamiltonian is given by
\begin{equation}
    E = E_B + E_T + E_V + E_A,
\end{equation}
where $E_B$ is the bending energy, $E_T$ is the tethering potential that constrains the length of the bonds in the network, and $E_V$ and $E_A$ penalize changes in the volume enclosed by the membrane and the membrane area, respectively. Specifically, $E_A$ and $E_V$ are given by
\begin{equation}
    E_A = \kappa_A \left( \frac{A-A_0}{A_0}\right)^2 \qquad E_V = \kappa_V \left( \frac{V-V_0}{V_0}\right)^2,
\end{equation}
where $A_0$ and $V_0$, which are set by the initial condition of the simulation, correspond to the area and volume of a perfect sphere with radius $R$. See Ref.~\cite{Siggel2022} for the specific  discretization of the bending energy $E_B$ and the functional form of the tethering potential $E_T$.
The constants $\kappa_B$, $\kappa_T$, $\kappa_A$ and $\kappa_V$ set the scale of the bending, tethering, area and volume energy contributions, respectively. 
We choose $\kappa_B = 20 ~k_B T$, where $k_B T$ is the simulation unit of energy, $\kappa_T = 10^{4} ~k_B T$ and deviations from $A_0$ and $V_0$ are penalised with $\kappa_{A,V} = 2.5\times 10^5 ~k_B T$ throughout the paper. 
We prevent the self-intersection of the triangulated network by including a tabulated soft repulsion between vertices beyond third neighbours in the lattice instead of using the repulsion potential from TriMem.
The simulation unit of length $\sigma$ corresponds to the lower bound at which the compression of the bonds in the network is penalised by the tethering potential $E_T$; the stretching of the bonds is penalised beyond $r = \sqrt{3}\sigma$.
We do not tether the center of mass of the vesicle in any of the simulations conducted. 
Throughout the paper we use a triangulated network with 5072 vertices, which leads to a vesicle with an initial diameter of $\sigma_{h} \approx 20 \sigma$.

\subsection{Non-reciprocal interactions in LAMMPS}
To achieve non-reciprocal interactions in simulation, we have developed a LAMMPS pair style that explicitly breaks action-reaction symmetry. 
This is realised by updating the forces acting on each pair of interacting particles with different contributions, according to Eq. (1) in the main text. The pair style is available under the name 'nonreciprocal' in the repository that accompanies the manuscript~\cite{TriLMP}, and it can be used as a pair style in LAMMPS by compiling the software as usual.

\subsection{Simulation dynamics}
To integrate the dynamics of the fluid lipid membrane, we perform Hybrid Monte Carlo (HMC) simulations. 
These simulations consist of two stages: a Molecular Dynamics stage (MD) and a Monte Carlo (MC) stage. 
In particular, a typical TriLMP run consists of a sequence of randomly alternating MD and MC stages. 
During an MD stage of the simulation, we time-integrate the dynamics of the membrane beads and symbiont particles in the NVT ensemble using the implementation of the Langevin thermostat available in LAMMPS. 
All particles in the system have mass $m = 1$ (mass scale of the system). 
The Langevin thermostat is characterised by temperature $k_B T = 1$ (energy scale of the system) and damping coefficient $\tau_D = 1~\tau$, where $\tau = \sigma \sqrt{m/k_B T}$ is the simulation unit of time. 
The damping coefficient of symbiont particles, with diameter $\sigma_s > \sigma$, is rescaled by $m/\sigma_s$ to correctly account for the friction acting on them by the implicit solvent. 
We use a time step $dt = 10^{-3}$ to integrate the equations of motion.
The length of each MD stage is set to 50 simulation steps.
During an MC stage of the simulation, all particles remain fixed in space, and we update the connectivity of the vesicle network by attempting to swap a maximum fraction $f$ of pairs of bonds according to a Metropolis-MC criterion~\cite{Siggel2022}. 
We choose $f = 0.2$, which ensures sufficient fluidity without compromising computational efficiency. 
We note that despite using a HMC scheme, we accept all configurations generated during the MD stage of the simulation as the system is intrinsically out-of-equilibrium due to the non-reciprocity of the interactions. 

\section{Other exponents for the non-reciprocal force}

\begin{figure*}

    \centering
    \includegraphics[width=\linewidth]{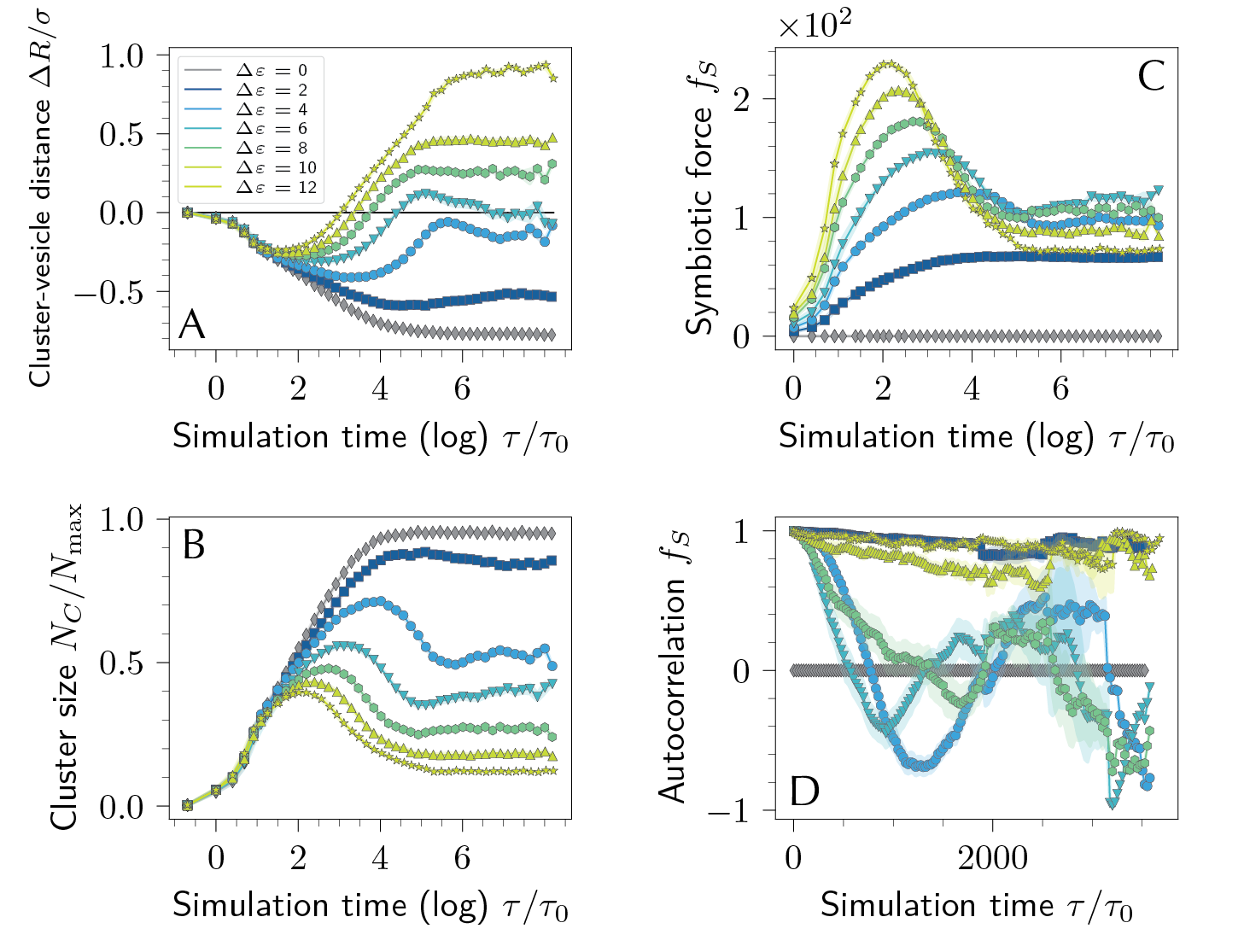}
    \caption{\textbf{Changing the exponent $\gamma$ in the symbiotic force}. Characterization of the host-symbiont active cluster for $\gamma = 2$. We plot the same observables shown in Fig.~3 in the main text. (A) Distance between the Center of Mass (COM) of the host-symbiont cluster and the COM of the host vesicle, as a function of time and for various activities $\Delta \varepsilon$. (B) Number of membrane beads within the host-symbiont cluster as a function of time, normalized by the equilibrium number of beads $N_{\max}~(\Delta \varepsilon = 0)$. (C) Symbiotic force as a function of time, and its autocorrelation function (D).}
    \label{fig:exponent}
\end{figure*}

We have tested that the specific functional form of the non-reciprocal pairwise force in Eq.~(1) does not alter the results qualitatively by simulating the system for two exponents, $\gamma = 7$ (results shown in the main text) and $\gamma = 2$, which corresponds to the exponent for phoretic interactions between catalytically-coated colloids in the far-field approximation~\cite{Soto2014}. In SI Fig.~\ref{fig:exponent}, we reproduce panel C in Fig.~3 in the main text for $\gamma=2$, where we plot the distance between the cluster and the center of mass of the vesicle, the host-symbiont cluster size, the non-reciprocal force and its autocorrelation function. The curves for $\Delta \varepsilon>0$ are non-monotonic, as described in the main text. Notice that while for weak activity $\Delta \varepsilon/k_B T = 2$ the symbiotic force does not decorrelate in this case (unlike for $\gamma = 7$), the size of the active cluster remains close to its equilibrium value.

\section{Morphology identification criteria}
Below we detail the criteria used to guide and identify the shapes in Figs.~2, 4 and 5. The analysis is conducted over the entire simulation trajectory, and shapes are reported if they are generated transiently during the simulation.

\subsection{Protrusion identification}
We define as a protrusion the collection of points that forms a tube that is longer than its base, and that has an acylindricity below a prescribed cutoff. The procedure for the identification of protrusions (Figs. 2 and 4) detailed below is inspired by the approach introduced by Iyer \textit{et al.}~\cite{Iyer2022}. 

Given a simulation snapshot, we start by computing the distribution of distances to the Center Of Mass (COM) of the vesicle for all the vertices in the membrane. 
We next select all vertices in the distributionbeyond a cutoff $R_{c} = r_{p} + d$ from the COM, where $r_{p}$ is the position of the peak in the distribution and $d = 5\sigma$ is chosen heuristically. 
This selects an ensemble of vertices, and generates protrusion candidates.
Nonetheless, as these vertices can constitute a catenoidal deformation, or correspond to deviations of the vesicle away from an average spherical shape, additional analysis steps must be conducted.

Since protrusions emerge from the interaction between the membrane and symbiont particles (Figs.~2 and 4), we only consider configurations where symbionts are within interaction range of the membrane. 
Consequently, vertices in a protrusion should be located between a symbiont particle and $R_c$, which allows us to better define the protrusion candidate.

The above procedure results in a collection of vertices that define a structure in space.  
To determine whether such structure corresponds to a protrusion, we compute the COM of its base, that is, the side closest to the vesicle COM, and use the vector that connects the base COM and the symbiont particle to align the candidate structure along the $z$ axis. This allows us to estimate the length as well as the mean radius of the candidate structure.
All structures with a base diameter larger than their length are discarded.
Finally, we compute the gyration tensor of the candidate and extract its acylindricity $a$.  
A candidate structure is considered a protrusion if its acylindricity is $a \leq 1$, which is also chosen heuristically. 

To identify the protrusion region in Fig. 2, we compute the probability of observing a protrusion in a sample of 10 simulation replicas. Each point in the protrusion region represents a probability $p\geq30\%$ of observing a protrusion. Consequently, the catenoid region is defined by all other points for which $p<30\%$.

\subsection{Budding identification}

\begin{figure*}
    \centering
    \includegraphics[width=\linewidth]{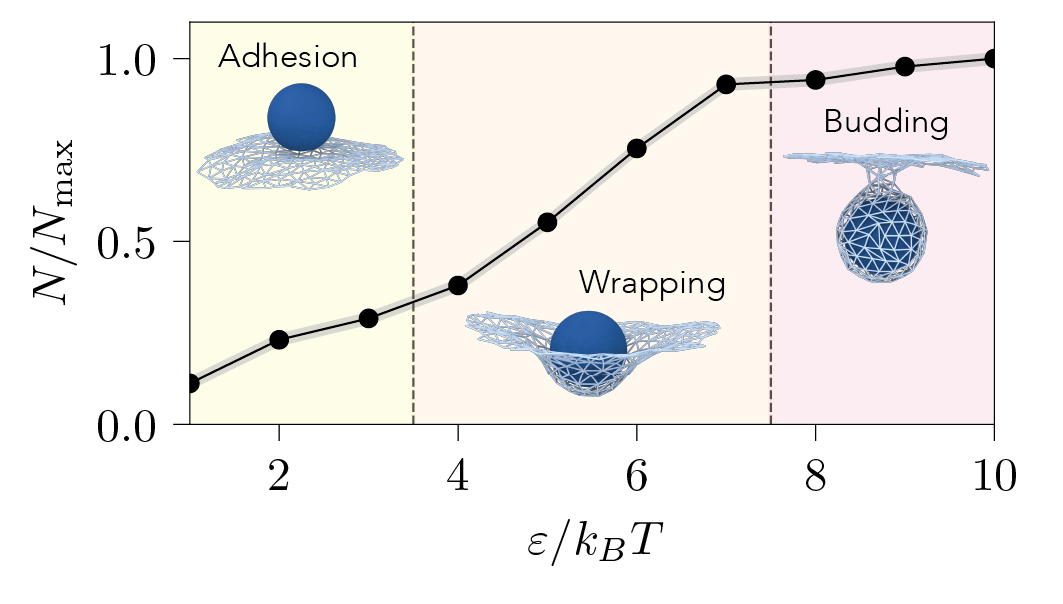}
    \caption{\textbf{Symbiont particle wrapping.} Normalized number of neighbouring membrane beads $N$ at equilibrium (light blue, membrane shown as network) for a symbiont particle (dark blue) as a function of the reciprocal membrane-symbiont interaction strength $\varepsilon$. The budding region is defined by the strengths $\varepsilon$ for which the number of neighbouring membrane beads $N/N_{\max}>0.9$.}
    \label{fig:reciprocalwrapping}
\end{figure*}

As we are using a triangulated membrane model with fixed topology, symbiont particles cannot fully bud into the vesicle. To determine the budding of a symbiont, we monitor the number of membrane beads $N$ within interaction range of a symbiont, which saturates upon symbiont budding (SI Fig.~\ref{fig:reciprocalwrapping}). We consider that the symbiont particle has budded if $N/N_{\max}\leq 0.9$, where $N_{\max}$ is the maximum number of neighbouring membrane beads measured for a symbiont particle with diameter $\sigma_s$. Likewise, we define wrapping when $0.3 \leq N/N_{\max} < 0.9$. For $N/N_{\max}<0.3$, the symbiont particle still remains adhered to the membrane but it does not deform it significantly. 

\subsection{Invagination identification}

Many of the invaginations observed in the simulations are transient, and eventually result in the symbiont particle being (doubly) wrapped by the membrane (see Supplementary Movie 3). 
This means that analysing the final snapshot of the simulation can hide the formation of an invagination in earlier stages. 
To distinguish the regions of the parameter space where invaginations form from those where the symbiont particle directly buds, we compute the distribution of distances to the COM of the vesicle for all the vertices in the membrane.
We then compute the maximum difference reached in simulation between the peak of such distribution and the position of the membrane-symbiont cluster with respect to the COM of the vesicle. 
We consider that an invagination has occurred if the maximum difference $\Delta \leq -5$ at some point in the simulation, indicating that the cluster formed by the membrane beads wrapped around the symbiont particle was deep within the vesicle. 

\section{Simulations with one-particle-thick model}\label{yuansection}

To show the independence of our results from the chosen membrane model, we also performed simulations with a coarse-grained, solvent free, one-particle-thick membrane model introduced by Yuan \textit{et al.}~\cite{Yuan2010}. This model consists of particles that interact via an anisotropic pair potential that drives the self-assembly of fluid membranes with bending rigidity $\kappa_B \approx 20~ k_B T$ for parameters $\mu = 3 $, $\zeta = 4.5$, $\theta_0 = 0$ and $\varepsilon_y = 5.8 ~k_B T$, and interaction range $r_{c, y} = 2.6~\sigma$, where $\sigma$ is the diameter of each particle in the membrane. 

Using this model, we observe the same morphologies as those reported in Fig. 2 (see Supplementary Fig.~\ref{fig:morphologyylz}). Nonetheless, we also find new configurations related to membrane rupture and pore formation which are not accessible to the triangulated membrane model with fixed topology. The one-particle-thick membrane can easily rupture during the formation of a protrusion. Likewise, the symbiont can get inserted in the membrane opening a pore when $\Delta \varepsilon \ll 0$ or bud after a tunnelling invagination. 
These points yield $N/N_{\max}>1$ for the triangulated membrane, i.e., they result in configurations where the symbiont is doubly-wrapped by the membrane; we represent them as white data points in Fig.~2 in the main text.

\section{Supplementary Videos}

\textbf{Supplementary Movie 1: }
Extrusion of a membrane protrusion as a result of the non-reciprocal interaction between the host vesicle and a single symbiont particle with diameter $\sigma_S/\sigma = 5$, for $\Delta \varepsilon/k_B T = \varepsilon_{hs}/k_BT = 10 $. When organisms unbind and stop interacting, the protrusion retracts. We explicitly plot the trajectory of the symbiont particle in time (from blue to red), which highlights its persistence as part of the active cluster during the interaction with the host vesicle. This contrasts the diffusive motion that the symbiont exhibits when host and symbiont are not interacting.

\textbf{Supplementary Movie 2:}
Transition from catenoid-like deformations to protrusion extrusions for increasing $\Delta \varepsilon$, with $ \varepsilon_{hs} = 9 $ and $\sigma_S/\sigma = 5$. From left to right, $\Delta \varepsilon/k_B T =1$, $\Delta \varepsilon/k_B T =3$, $\Delta \varepsilon/k_B T =5$, $\Delta \varepsilon/k_B T =7$ and $\Delta \varepsilon/k_B T =9$. The simulation is oriented such that the vector that connects the center of mass of the vesicle and the symbiont particle is aligned with the z-axis in each frame to aid in the visualization. The center of mass of the vesicles is fixed at (0, 0, 0) for visualization, and we additionally show the triad of unit vectors in the lab frame for reference (bottom left of the vesicle for each simulation). 

\textbf{Supplementary Movie 3:}
Morphologies reported as symbiont invagination (left) and budding (right), for $\Delta \varepsilon/k_B T = -2$ and $\varepsilon_{sh}/k_BT = 1$ (left) and $\varepsilon_{sh}/k_BT = 9$ (right). The center of mass of the vesicles is fixed at (0, 0, 0) in the movie. For each snapshot, we calculate the centroid of the membrane beads within range $r < 5\sigma$ from the symbiont particle, and align the vector that connects the symbiont and the centroid with the z-axis to aid in the visualization of the morphology. This alignment procedure, which only affects the visualization, explains the initial snapshots of the movie. The triad of unit vectors in the lab frame is also shown for reference (bottom left of the vesicle for each simulation). 

\textbf{Supplementary Movie 4: }
Cyclic formation of membrane blebs, for a host vesicle interacting with $N_S = 5$ symbionts, $\Delta \varepsilon/k_B T = -2$ and $\varepsilon_{sh}/k_B T = 2$. The host vesicle is rendered with increased transparency to facilitate the visualization of the symbiont particles. The center of mass of the vesicles is fixed at (0, 0, 0) for visualization.

\textbf{Supplementary Movie 5: }
Ballistic motion of a host vesicle as a result of its interaction with $N_S = 5$ symbionts for $\Delta \varepsilon/k_BT = -6$ and $\varepsilon_{sh}/k_BT = 2$. The vesicle is shown in its frame of reference (left) and moving in the simulation box (right). The trajectory of the center of mass of the vesicle is explicitly shown in time.

\textbf{Supplementary Movie 6: }
Extrusion of protrusions for $N_S = 10$ symbionts, $\varepsilon_{sh} = 2~k_B T$ and $\Delta \varepsilon = -4~k_B T$. The center of mass of the vesicle is pinned for visualization.

\textbf{Supplementary Movie 7: }
Extrusion of protrusions by for $N_S = 50$ symbionts, $\varepsilon_{sh} = 2~k_B T$ and $\Delta \varepsilon = -2~k_B T$.

\begin{figure*}
    \centering
    \includegraphics[width=\linewidth]{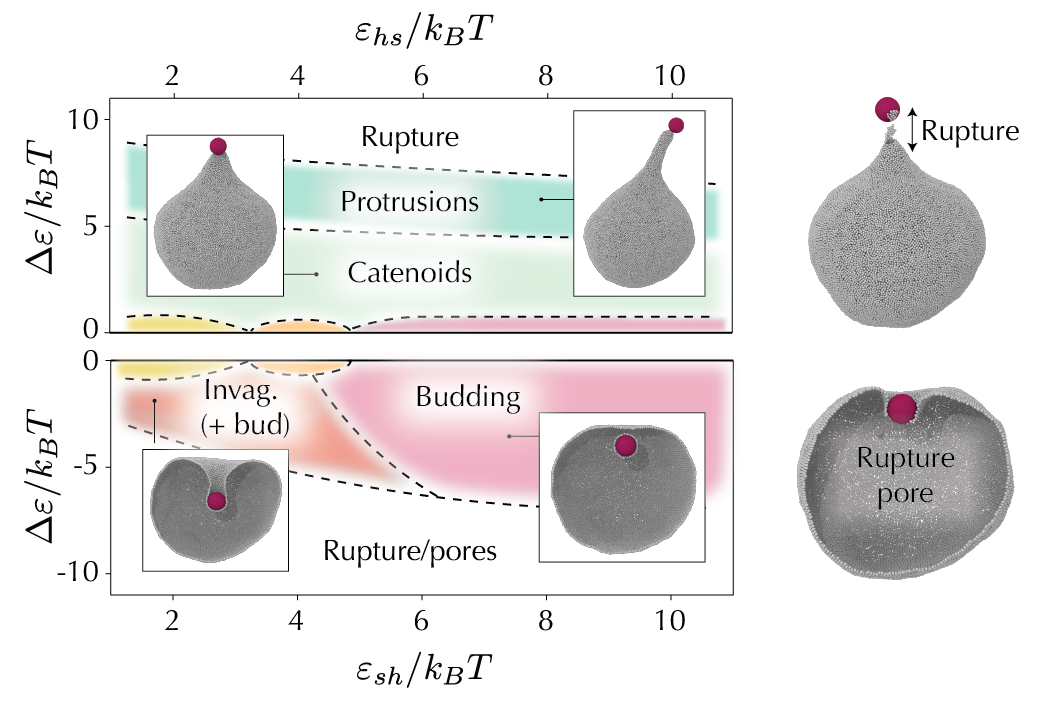}
    \caption{\textbf{Deformations of a one-particle thick membrane}. Morphology diagram for a symbiont particle (pink) interacting with a one-particle-thick fluid membrane model (gray). The diagram obtained for this membrane model recapitulates the morphologies described in Fig.~2 in the main text for a triangulated membrane model. Nonetheless, as the one-particle-thick membrane can undergo topological changes, the membrane often breaks upon encountering the symbiont at high activity, i.e., $\Delta \varepsilon \ll 0$ and $\Delta \varepsilon \gg 0$. Data corresponds to MD simulations with a membrane with $N = 15872$ particles, and $\sigma_s = 10~\sigma$.}
    \label{fig:morphologyylz}
\end{figure*}

\begin{figure*}
    \centering
    \includegraphics[width=\linewidth]{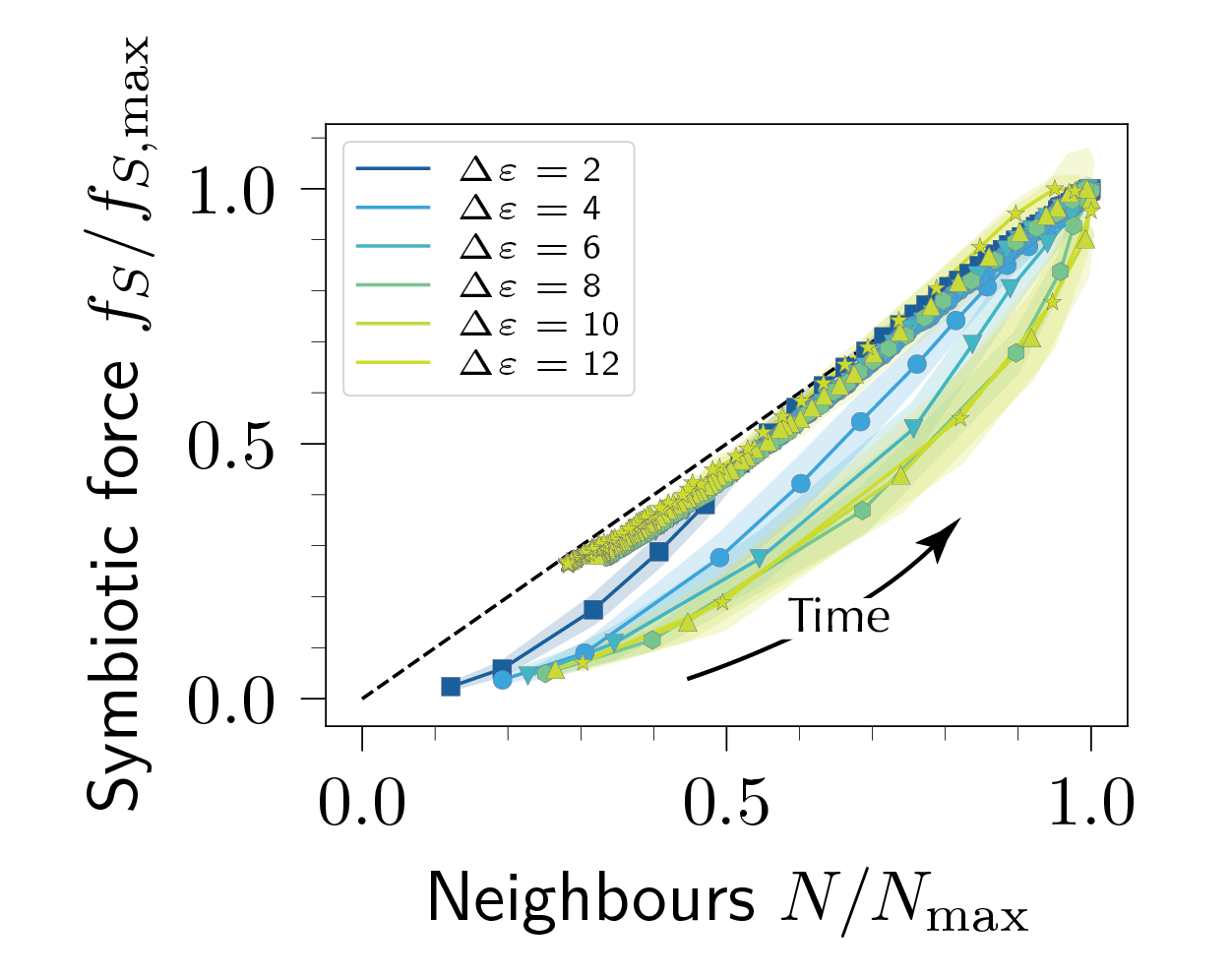}
    \caption{\textbf{Correlation between the symbiotic force and the active cluster size.} Symbiotic force as a function ofthe number of neighbours within the host-symbiont active cluster, for various activities $\Delta \varepsilon$. Data corresponds to different simulation time points, as indicated by the black arrow on the left. By following the arrow, we recover the trajectory of the cluster.}
    \label{fig:correlation}
\end{figure*}

\begin{figure*}
    \centering
    \includegraphics[width=\linewidth]{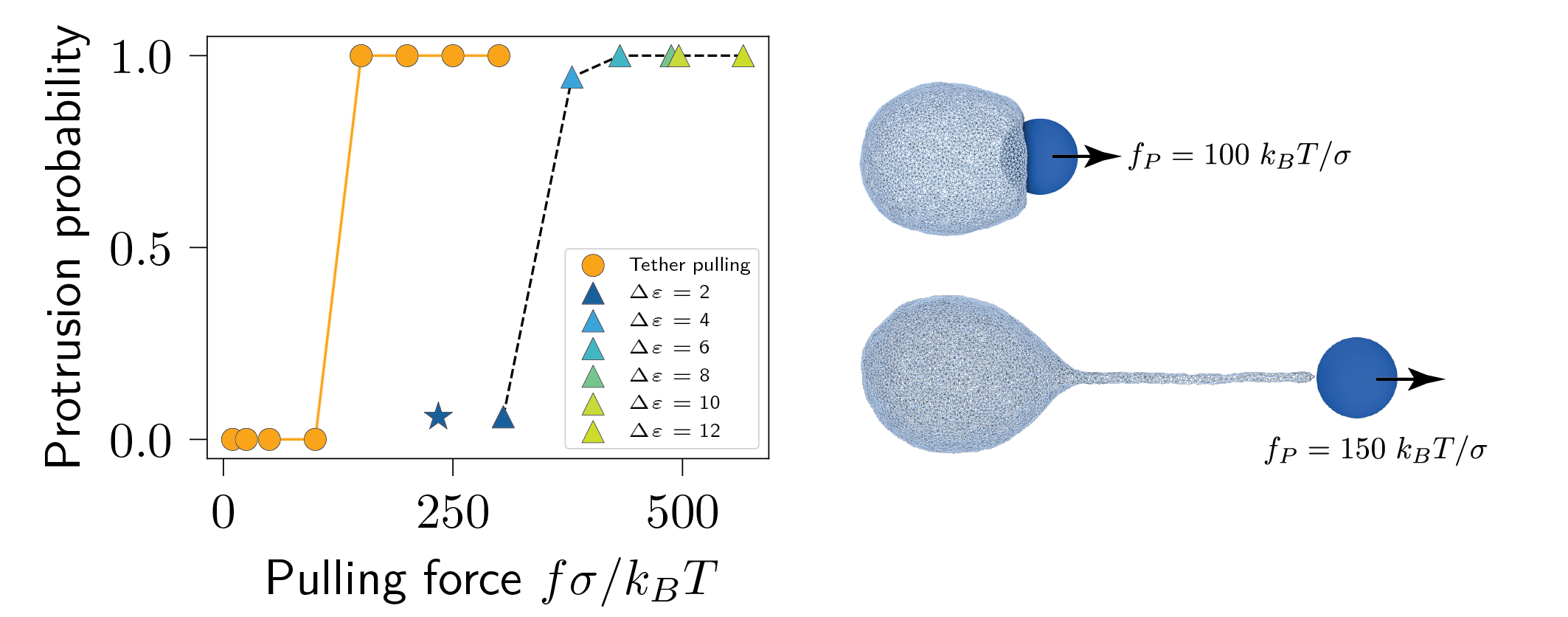}
    \caption{\textbf{Comparison between reciprocal and non-reciprocal pulling}. Protrusion extrusion probability as a function of the pulling force. Orange circles (average over 3 simulation replicas) correspond to simulations where the symbiont particle ($\sigma_s = 20\sigma$) is tethered to one membrane bead and allowed to interact with the membrane reciprocally with $\varepsilon_{hs} = \varepsilon_{sh} = 10~k_B T$; a force $f_P$ is then applied to the symbiont particle to extrude the protrusion (right simulation snapshots). Triangles (average over at least 10 simulation replicas) correspond to the non-reciprocal interaction between the host and the symbiont particle, with varying $\Delta \varepsilon$. The pulling force corresponds to the symbiotic force in this case. We plot the maximum force achieved upon interaction between organisms (see Fig.~3 in the main text). The dark blue star corresponds to the steady-state force for $\Delta \varepsilon = 2 ~k_B T$.}
    \label{fig:pulling}
\end{figure*}

\begin{figure*}
    \centering
    \includegraphics[width=\linewidth]{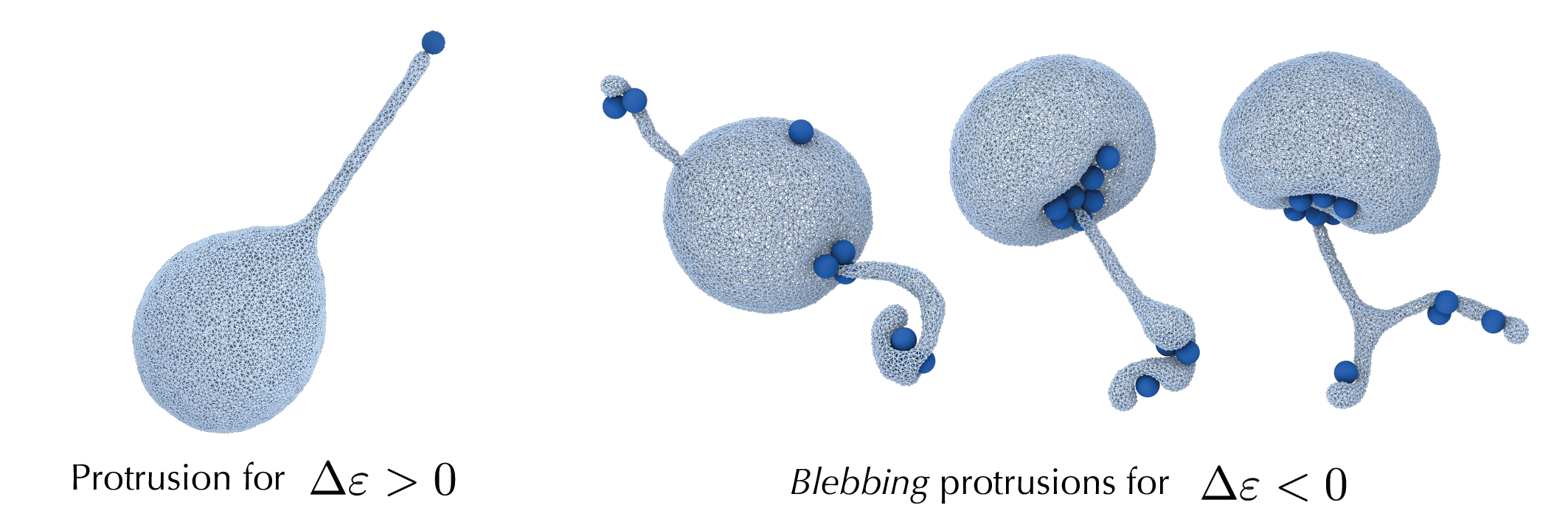}
    \caption{\textbf{Comparison of protrusions for $\Delta\varepsilon>0$ (left) and $\Delta\varepsilon<0$ (right)}. Simulation snapshot on the left corresponds to $N_S = 1$, $\varepsilon_{hs} = 10~k_B T$ and $\Delta \varepsilon = 10~k_B T$. Simulation snapshots on the right correspond to $N_S = 10$, $\varepsilon_{sh} = 2~k_B T$ and $\Delta \varepsilon = -4~k_B T$. The dynamics of the system can be visualized in SI Movie 6.}
    \label{fig:chubbyprot}
\end{figure*}

\begin{figure*}
    \centering
    \includegraphics[width=\linewidth]{SIFig7.png}
    \caption{\textbf{Steady-state morphologies for model cell interacting with multiple symbionts.} Data corresponds to $\varepsilon_{hs}/k_B T = 2.$; three replicas are shown per parameter.}
    \label{fig:finalsnap}
\end{figure*}

%\bibliography{bibliography}

\end{document}